\documentclass[twocolumn,aps,showpacs,superscriptaddress]{revtex4}
\usepackage{graphicx}
\usepackage{psfig}

\newcommand{ \be }{\begin{equation}}
\newcommand{ \ee }{\end{equation}}
\newcommand{ \bea }{\begin{eqnarray}}
\newcommand{ \eea }{\end{eqnarray}}
\newcommand{ \la }{\langle}
\newcommand{ \ra }{\rangle}

\newcommand{\p}{$\phi$}

\newcommand{\energy}{$\sqrt{s_{_{NN}}}=200$~GeV}
\begin{document}
\title{

$\phi$ meson production in $Au+Au$ and $p+p$ collisions at
$\sqrt{s_{_{NN}}}=200$~GeV }

%\documentclass[aps,prl,superscriptaddress]{revtex4}
%\begin{document}
%\title{STAR Collaboration Author List - November 7, 2003}
%STAR author list in APS format for use with revtex4 i.e. for PRL, PRC etc.
%Force affiliation order to be alphabetic, otherwise they come out in the author order

\affiliation{Argonne National Laboratory, Argonne, Illinois 60439}
\affiliation{Brookhaven National Laboratory, Upton, New York 11973}
\affiliation{University of Birmingham, Birmingham, United Kingdom}
\affiliation{University of California, Berkeley, California 94720}
\affiliation{University of California, Davis, California 95616}
\affiliation{University of California, Los Angeles, California 90095}
\affiliation{California Institute of Technology, Pasadena, California 91125}
\affiliation{Carnegie Mellon University, Pittsburgh, Pennsylvania 15213}
\affiliation{Creighton University, Omaha, Nebraska 68178}
\affiliation{Nuclear Physics Institute AS CR, \v{R}e\v{z}/Prague, Czech Republic}
\affiliation{Laboratory for High Energy (JINR), Dubna, Russia}
\affiliation{Particle Physics Laboratory (JINR), Dubna, Russia}
\affiliation{University of Frankfurt, Frankfurt, Germany}
\affiliation{Indian Institute of Technology, Mumbai, India}
\affiliation{Indiana University, Bloomington, Indiana 47408}
\affiliation{Institute  of Physics, Bhubaneswar 751005, India}
\affiliation{Institut de Recherches Subatomiques, Strasbourg, France}
\affiliation{University of Jammu, Jammu 180001, India}
\affiliation{Kent State University, Kent, Ohio 44242}
\affiliation{Lawrence Berkeley National Laboratory, Berkeley, California 94720}\affiliation{Max-Planck-Institut f\"ur Physik, Munich, Germany}
\affiliation{Michigan State University, East Lansing, Michigan 48824}
\affiliation{Moscow Engineering Physics Institute, Moscow Russia}
\affiliation{City College of New York, New York City, New York 10031}
\affiliation{NIKHEF, Amsterdam, The Netherlands}
\affiliation{Ohio State University, Columbus, Ohio 43210}
\affiliation{Panjab University, Chandigarh 160014, India}
\affiliation{Pennsylvania State University, University Park, Pennsylvania 16802}
\affiliation{Institute of High Energy Physics, Protvino, Russia}
\affiliation{Purdue University, West Lafayette, Indiana 47907}
\affiliation{University of Rajasthan, Jaipur 302004, India}
\affiliation{Rice University, Houston, Texas 77251}
\affiliation{Universidade de Sao Paulo, Sao Paulo, Brazil}
\affiliation{University of Science \& Technology of China, Anhui 230027, China}
\affiliation{Shanghai Institute of Nuclear Research, Shanghai 201800, P.R. China}
\affiliation{SUBATECH, Nantes, France}
\affiliation{Texas A\&M University, College Station, Texas 77843}
\affiliation{University of Texas, Austin, Texas 78712}
\affiliation{Valparaiso University, Valparaiso, Indiana 46383}
\affiliation{Variable Energy Cyclotron Centre, Kolkata 700064, India}
\affiliation{Warsaw University of Technology, Warsaw, Poland}
\affiliation{University of Washington, Seattle, Washington 98195}
\affiliation{Wayne State University, Detroit, Michigan 48201}
\affiliation{Institute of Particle Physics, CCNU (HZNU), Wuhan, 430079 China}
\affiliation{Yale University, New Haven, Connecticut 06520}
\affiliation{University of Zagreb, Zagreb, HR-10002, Croatia}
%Authors 1 per line follow by affiliation, three authors have two affiliations
\author{J.~Adams}\affiliation{University of Birmingham, Birmingham, United Kingdom}
\author{C.~Adler}\affiliation{University of Frankfurt, Frankfurt, Germany}
\author{M.M.~Aggarwal}\affiliation{Panjab University, Chandigarh 160014, India}
\author{Z.~Ahammed}\affiliation{Variable Energy Cyclotron Centre, Kolkata 700064, India}
\author{J.~Amonett}\affiliation{Kent State University, Kent, Ohio 44242}
\author{B.D.~Anderson}\affiliation{Kent State University, Kent, Ohio 44242}
\author{D.~Arkhipkin}\affiliation{Particle Physics Laboratory (JINR), Dubna, Russia}
\author{G.S.~Averichev}\affiliation{Laboratory for High Energy (JINR), Dubna, Russia}
\author{S.K.~Badyal}\affiliation{University of Jammu, Jammu 180001, India}
\author{J.~Balewski}\affiliation{Indiana University, Bloomington, Indiana 47408}
% ...

%\collaboration{STAR Collaboration}\homepage{www.star.bnl.gov}\noaffiliation

\collaboration{STAR Collaboration}

%\maketitle
%\end{document})

%--====================================================
\begin{abstract}

We report the STAR measurement of $\phi$ meson production in
$Au+Au$ and $p+p$ collisions at $\sqrt{s_{_{NN}}}=200$~GeV. Using
the event mixing technique, the $\phi$ spectra and yields are
obtained at mid-rapidity for five centrality bins in $Au+Au$
collisions and for non-singly-diffractive $p+p$ collisions. It is
found that the $\phi$ transverse momentum distributions from
$Au+Au$ collisions are better fitted with a single-exponential
while the $p+p$ spectrum is better described by a
double-exponential distribution. The measured nuclear modification
factors indicate that $\phi$ production in central $Au+Au$
collisions is suppressed relative to peripheral collisions when
scaled by the number of binary collisions ($\la N_{bin} \ra$). The
systematics of $\langle p_t \rangle$ versus centrality and the
constant $\phi/K^-$ ratio versus beam species, centrality, and
collision energy rule out kaon coalescence as the dominant
mechanism for $\phi$ production.

\end{abstract} 

\maketitle
%----------------------------------------------------------------------

In elementary collisions the production of the $\phi$ meson, the
lightest bound state of strange quarks ($s\bar{s}$), is suppressed
because of the OZI rule \cite{okubo,zweig,iizuka}. In heavy-ion
collisions, however, strange quarks are produced copiously and
$\phi$ enhancement is observed relative to expectations from $p+p$
collisions \cite{agsphi,na49phi,phiy1}. Theoretical calculations
have tried to address the origins of this enhancement
\cite{dover,shor,sorge}.  The $\phi$ meson is also thought to have
a small hadronic cross-section \cite{perkins} and may provide
direct information about the dense matter at hadron formation
without perturbations from co-moving hadrons. For these reasons,
$\phi$ production in relativistic nuclear collisions has been of
great interest.

The mechanism for $\phi$ production in high energy collisions has
remained an open issue.  A naive interpretation of the $\phi$
enhancement observed in heavy-ion collisions would be that the
$\phi$ is produced hadronically via $K \bar{K} \rightarrow \phi$.
Hadronic rescattering models such as RQMD and UrQMD
\cite{urqmd,RQMD}, implementing such processes, predict an
increase in the $\phi/K^-$ ratio as a function of the number of
participants. Rescattering models also predict similar increases
in the $\langle p_t \rangle$ of the proton and $\phi$ meson.

The nuclear modification factors ($R_{AA}$ and $R_{CP}$) of the
$\phi$ meson are important in differentiating between mass and
particle species ordering. Current measurements of identified
hadrons by STAR ($\Lambda$ and $K^0_S$) show that $R_{CP}$ for the
$\Lambda$ differs from that of the $K^0_S$ \cite{laksstar}. It is
difficult, however, to determine whether this difference is
related to the mass of the particle or the type of the particle
(whether it is a baryon or a meson) since there is a significant
mass difference between the $\Lambda$ and the $K^0_S$. The $\phi$,
however, has a mass that is similar to that of the $\Lambda$, yet
is a meson. A direct comparison of the $\phi$ $R_{CP}$ and
$R_{AA}$ with these previous measurements will provide more
insight into this mass vs. particle species dependence.

The STAR detector \cite{STARdet} consists of several sub-systems
in a large solenoidal analyzing magnet. For the data taken during
the second RHIC run (2001-2002) presented here, the experimental
setup consisted of a Time Projection Chamber (TPC), a Central
Trigger Barrel (CTB), a pair of Beam-Beam Counters (BBC), and two
Zero Degree Calorimeters (ZDC). The ZDC's are used as the
experimental trigger for $Au+Au$ collisions while the BBC's are
used for the $p+p$ trigger.

The results presented here were obtained from about 2.1 million
minimum-bias $Au+Au$ events, 1.1 million central $Au+Au$ events
and 6.5 million non-singly-diffractive (NSD) $p+p$ events.
Reconstruction of the $\phi$ was accomplished by calculating the
invariant mass ($m_{inv}$), transverse momentum ($p_t$), and
rapidity ($y$) of pairs that formed from all permutations of
candidate $K^+$ with $K^-$.  The resulting $m_{inv}$ distribution
consisted of the $\phi$ signal atop a large background that is
predominantly combinatorial.  The shape of the combinatorial
background was calculated using the mixed-event technique
\cite{mixing1,mixing2}.

For the centrality measurement, the raw hadron multiplicity
distribution within a pseudo-rapidity window $| \eta | \le$ 0.5 is
divided into five bins corresponding to 50--80\%, 30-50\%,
10--30\%, top 10\% and top 5\% of the measured cross section for
$Au+Au$ collisions. Events are selected with a primary vertex $z$
position from the center of the TPC of $|z| < 25$~cm for $Au+Au$
collisions and $|z| < 50$~cm for $p+p$ collisions, where $z$ is
along the beam axis. These events are further divided according to
$z$ to reduce acceptance-induced distortions in the mixed-event
background. Correlations in the background due to elliptic flow
were minimized by mixing events with similar reaction plane
angles. Consistent results are obtained when we construct the
background distribution using like-sign pairs from the same-event.

Particle identification (PID) is achieved by correlating the
ionization energy loss ($dE/dx$) of charged particles in the TPC
gas with their measured momentum.
The measured $\la dE/dx \ra$ is reasonably well described by the
Bethe-Bloch function \cite{perkins,phiee} smeared with a resolution of
width $\sigma$.  Tracks within $2\sigma$ of the kaon Bethe-Bloch
curve are selected for this analysis.

\begin{figure}[h]
\centering\mbox{\psfig{figure=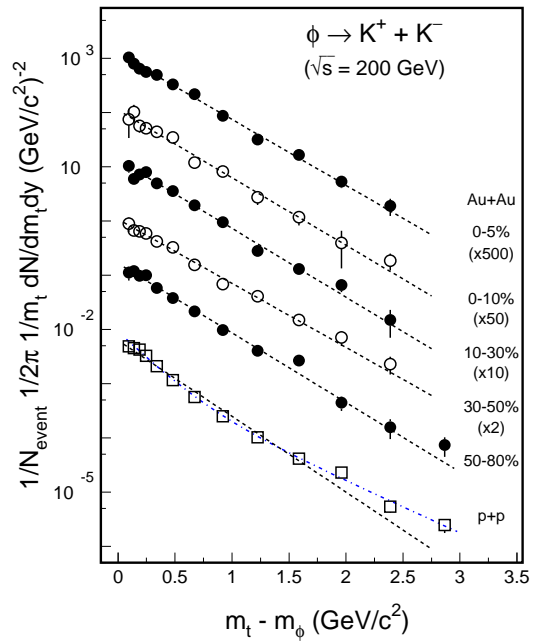,width=.45\textwidth}}
\vspace{-0.75cm} \caption{ The transverse mass distributions from
$Au+Au$ (circles) and $p+p$ (squares) collisions at $200$~GeV. For
clarity, some $Au+Au$ distributions for different centralities are
scaled by factors. The top 5\% data are obtained from the central
trigger data set. All other distributions are obtained from the
minimum-bias data set. Dashed lines represent the exponential fits
to the distributions and the dotted-dashed line is the result of
a double-exponential fit to the distribution from $p+p$ collisions. Error
bars are statistical errors only.} \label{dndpt}
\end{figure}

To obtain the $\phi$ spectra, same-event and mixed-event
distributions are accumulated and background subtraction is done
in each $p_t$, $y$ and centrality bin.  The mixed-event background
$m_{inv}$ distribution is normalized to the same-event $m_{inv}$
distribution in the region above the \p~ mass ($1.04 < m_{inv} <
1.2\;\mathrm{GeV/c^2}$).  A small, smooth residual background can
remain near the $\phi$ peak in the subtracted $m_{inv}$
distribution, because the mixed-event sample does not perfectly
account for the production of background pairs (kaons and/or pions
from PID leak-through) that are correlated, either by Coulomb or
other interactions or by such instrumental effects as track
merging \cite{phdthesis}. The raw yield in each bin is then
determined by fitting the background subtracted $m_{inv}$
distribution to a Breit-Wigner function plus a linear background
in a limited mass range. The measured mass and width of the $\phi$
are consistent with the value listed by the Particle Data Group
\cite{phiee} convoluted with detector resolution.

Using GEANT and detector response simulations, the data are
corrected for acceptance, kaon decay and tracking efficiencies to
obtain the final distributions presented here. Figure \ref{dndpt}
shows the transverse mass distributions from $Au+Au$ (circles) and
NSD $p+p$ (squares) collisions at 200 GeV. The spectra are obtained
from the rapidity range $|y_{\phi}| < 0.5$.  For clarity, some
$Au+Au$ distributions for different centralities are scaled by
factors indicated in the figure. Dashed lines represent
exponential fits to the distributions and the dotted-dashed line
represents a double-exponential fit to the $p+p$ result.

Statistical uncertainties are shown in the figure and the results
of the fits are listed in Table \ref{tab:vals}. The main
contributions to the systematic uncertainty come from fitting to
the $K^+K^-$ invariant-mass distribution, tracking and the PID
efficiency calculation. Different background functions and
normalization factors for the mixed-event background were used to
determine the uncertainty in the fitting to the invariant-mass
distribution and is estimated to be about 5\%. The uncertainty
from tracking and PID efficiency is estimated, by varying the
tracking and PID cuts on the daughter tracks, to be 8\%. The
overall systematic uncertainty in the yield, $dN/dy$ and $\langle
p_t \rangle$ is estimated to be 11\%, and includes an additional
contribution from fitting the transverse momentum distributions.
For $Au+Au$ collisions, the inverse slope parameters and yields
are extracted from a single exponential function fit. For $p+p$
collisions, however, there is an additional component beyond a
single exponential, see dashed-line in figure \ref{dndpt}. The
power-law shape provides a better fit at the higher $p_t$ region
but failed at low $p_t$. Double-exponential function provided a
better fit so it was used to extract the values of $dN/dy$ and
$\langle p_t \rangle$ for the $p+p$ collisions. For the heavy ion
results, a Boltzmann distribution and a thermal+flow model
\cite{blastw} are also used to fit the data as a check of the
systematic uncertainty in the extrapolated yield and $\langle p_t
\rangle$. The systematic uncertainty is $\sim$ 15\% in the overall
normalization and $\le$ 5\% in mean $p_t$ for the $p+p$ data,
including uncertainties in the vertex efficiency for very low
multiplicity events.

%%%%%%%%%%%%%%%%%%%%%%%%%%%%%%%%%%%%%%%
\begin{table}
\begin{tabular*}{\hsize}{lccc}
\hline \hline
{Centrality} & {Slope (MeV)} & {$\la p_t \ra$ (GeV/c)} & {$dN/dy$}\\
\hline
{0--5\%} & {$363 \pm 8$} & {$0.97 \pm 0.02$} & {$7.70 \pm 0.30$}\\
{0--10\%} & {$357 \pm 14$} & {$0.95 \pm 0.03$} & {$6.65 \pm 0.35$}\\
{10--30\%} & {$353 \pm 8$} & {$0.97 \pm 0.02$} & {$3.82 \pm 0.19$}\\
{30--50\%} & {$383 \pm 10$} & {$1.02 \pm 0.03$} & {$1.72 \pm 0.06$}\\
{50--80\%} & {$344 \pm 9$} & {$0.94 \pm 0.02$} & {$0.48 \pm 0.02$}\\
{$p+p$ minbias} & {$--$} & {$0.82 \pm 0.03$} & {$0.018 \pm 0.001$}\\
%{$p+p$ minbias} & {$266 \pm 9$} & {$0.75 \pm 0.02$} & {$0.020 \pm 0.001$}\\
\hline \hline
\end{tabular*}

\caption{Results of $\phi$ meson inverse slope parameter, $\langle p_t
\rangle$, and $dN/dy$ from NSD $p+p$ and $Au+Au$ collisions at RHIC. An
exponential is used for the $Au+Au$ data while a
double-exponential fit is used for the $p+p$ data. All values
are for $|y| < 0.5$ and only statistical errors are quoted.}
\label{tab:vals} \end{table}

%%%%%%%%%%%%%%%%%%%%%%%%%%%%%%%%%%%%%%%%

\begin{figure}%[tbh]
\centering\mbox{\psfig{figure=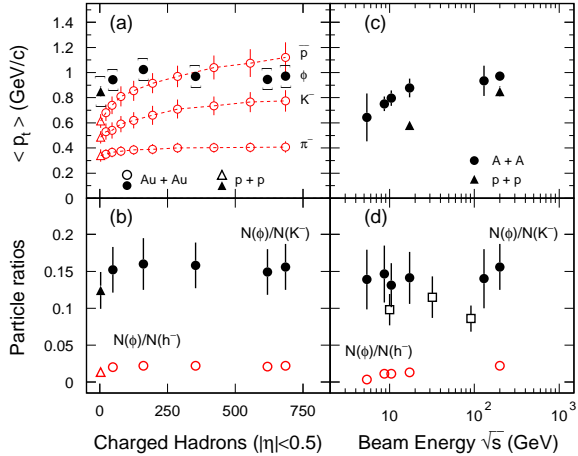,width=.49\textwidth}}
\caption{(a) $\phi$ $\langle p_t \rangle$ vs. measured number of
charged hadrons ($N_{ch}$) within $|\eta| \le 0.5$ at $200$~GeV.
For comparison, the values of $\langle p_t \rangle$ for negative
pions, kaons, and anti-protons are also shown; (b) Ratios of
N($\phi$)/N($K^-$), filled symbols, and N($\phi$)/N($h^-$), open
symbols, vs. $N_{ch}$; (c) $\langle p_t \rangle$ vs.
center-of-mass beam energy from central nucleus-nucleus (filled
circles) and $p+p$ collisions (filled triangles); (d) Ratios of
N($\phi$)/N($K^-$) from central nucleus-nucleus collisions, filled
circles, and N($\phi$)/N($h^-$), open circles, vs.  center-of-mass
beam energy. N($\phi$)/N($K^-$) ratio from $e^+e^-$ collisions
(open squares) are also shown. Note: All plots are from
mid-rapidity. Both the statistical and systematic errors are shown
for the 200 GeV STAR data, while only statistical errors are shown
for the energy dependence of the particle ratios.} \label{ratios}
\end{figure}

 The system-size and beam-energy dependence of $\langle p_t \rangle$,
 $\phi/K^-$ and $\phi/h^-$ are shown in figure \ref{ratios}. For
 comparison, the $\langle p_t \rangle$ of the $\bar{p}$, $K^-$ and
 $\pi^-$ are also shown \cite{200gevspectra}. In the $\phi/h^-$ ratio
 at \energy \ shows no significant dependence on centrality for $Au+Au$
 collisions, but decreases by about 30\% for $p+p$ collisions,
 see open symbols in plot (b). As a function of energy, see plots (c)
 and open
 circles in plot (d), both values of $\langle p_t \rangle$ and
 $\phi/h^-$ ratio increase. This indicates that the production of
 $\phi$ mesons is sensitive to the initial conditions of the collision.

The general trend for $\bar{p}$, $K^-$ and $\pi^-$ is an increase in
$\langle p_t \rangle$ as a function of centrality, which is indicative
of an increased transverse radial flow velocity component to these
particles' momentum distributions. The $\phi$ $\langle p_t \rangle$,
however, shows no significant centrality dependence. This indicates
that the $\phi$ does not participate in the transverse radial flow as
does the $\bar{p}$, $K^-$ and $\pi^-$. This is expected if the $\phi$
decouples early on in the collision before transverse radial flow is
completely built up. If the $\phi$ hadronic scattering cross section
is much smaller than that of other particles, one would not expect the
$\phi$ $\langle p_t \rangle$ distribution to be appreciably affected
by any final state hadronic rescatterings. In contrast to these
observations, the RQMD predictions of $\langle p_t \rangle$ for kaon,
proton and $\phi$ all increase as functions of centrality
\cite{RQMD,meanpt}.

The yield ratio $\phi/K^-$ from this analysis is constant as a
function of centrality and species ($p+p$ or $Au+Au$). In fact,
for collisions above the threshold for $\phi$ production, the
$\phi/K^-$ ratio is essentially independent of system size,
$e^+e^-$ to nucleus-nucleus, and energy from a few GeV up to 200
GeV (Figure \ref{ratios} (d))
\cite{agsphi,agsk,na49phi,na49kaon,phiy1,kyear1,phiee}. This is
remarkable, considering that the initial conditions of an $e^+e^-$
collision are so drastically different from $Au+Au$ collisions.
This observation may indicate that the ratio is dominated by the
hadronization process.

Rescattering models (RQMD \cite{RQMD}, UrQMD \cite{urqmd}) predict
that about 2/3 of $\phi$ mesons come from kaon coalescence in the
final state. The centrality dependence of the $\phi/K^-$ ratio
alone provides a serious test of the current rescattering models.
In these models, such as UrQMD, rescattering channels for $\phi$
production includes $K\bar{K}$ and $K$-Hyperon modes and predicts
an increasing $\phi/K^-$ ratio vs. centrality. These models also
predict an increase in $\la p_t \ra$ for the proton, kaon, and
$\phi$ of 40 to 50\% from peripheral to central collisions. A
comparison of the data to these models does not support the kaon
coalescence production mechanism for $\phi$ mesons.

\begin{figure}%[tbh]
\centering\mbox{\psfig{figure=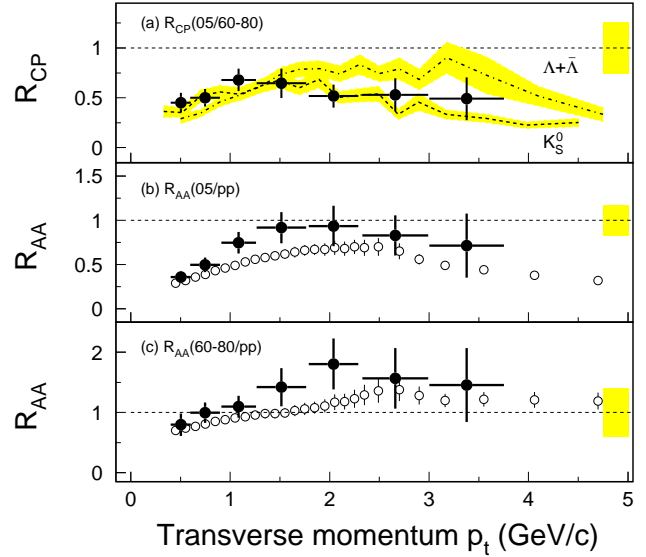,width=.49\textwidth}}

\caption{ $R_{CP}$ (a): The ratio of central (top 5\%) over
peripheral (60-80\%) ($R_{CP}$) normalized by $\la N_{bin} \ra$.
The ratios for the $\Lambda$ and $K^0_S$, shown by dotted-dashed
and dashed lines, are taken from \cite{laksstar}; $R_{AA}$ (b) and
(c) are the ratios of central $Au+Au$ (top 5\%) to $p+p$ and
peripheral $Au+Au$ (60-80\%) to $p+p$, respectively. The values of
$R_{AA}$ for charged hadrons are shown as open circles
\cite{highpt}. The width of the gray bands represent the
uncertainties in the estimation of $\la N_{bin} \ra$ summed in
quadrature with the normalization uncertainties of the spectra.
Errors on the $\phi$ data points are the statistical plus 15\%
systematic errors.} \label{raa}
\end{figure}

The particle-type dependence of the nuclear modification factors
$R_{AA}$ and $R_{CP}$ \cite{laksstar,highpt} should be sensitive
to the production dynamics and the hadronization process
\cite{zlin,voloshin,molnar,greco,fries}. $R_{AA}$ is the ratio of
the differential yield in a centrality class of $Au+Au$ collisions
to the inelastic differential cross-section in $p+p$ collisions,
scaled by the overlap integral $T_{AA} = \la N_{bin} \ra
/\sigma_{inel}$ from a Glauber calculation\cite{130pion}. The
Glauber calculation was performed with $\sigma_{inel} = 42 \pm
1$~mb. The inelastic differential cross-section in $p+p$ is
estimated as the NSD yield times $\sigma_{NSD}$, measured as $30.0
\pm 3.5$~mb, with a small correction, determined from Pythia
calculations, of 1.05 at pt=0.4~GeV/c and unity above 1.2~GeV/c
\cite{highpt}. $R_{CP}$ is the ratio of the yields between two
Au+Au centrality classes, scaled by $\la N_{bin} \ra$. The
$R_{CP}$ (Figure \ref{raa}(a)) for the $\phi$ meson at moderate
$p_t$ ($ 1.5 < p_t < 4$~GeV/c) is suppressed relative to the
binary collision scaling (dashed horizontal line at unity).

A comparison of the $R_{CP}$ for the $\phi$, $K^0_S$ and $\Lambda$
is shown in figure \ref{raa} (a). Both statistical and systematic
errors are included in the figure. The ratio $R_{AA}$ for central
(top 5\%) and peripheral (60-80\%) $Au+Au$ data are shown in
Figure \ref{raa} (b) and (c), respectively. $R_{AA}$ for charged
hadrons \cite{highpt} is also shown as a reference.  The charged
hadron and $\phi$ peripheral $R_{AA}$ both go above the binary
scaling limit, but are consistent with unity within the systematic
uncertainties.  The $\phi$ central $R_{AA}$ approaches unity and
point to point is higher than $R_{CP}$. With the systematic
uncertainty on the normalization of the ratio, however, both
$R_{AA}$ and $R_{CP}$ are consistent. Note that a $R_{AA}$ ratio
that is higher than the $R_{CP}$ ratio would be consistent with
OZI suppression of $\phi$ production in $p+p$
\cite{okubo,zweig,iizuka} and/or strangeness enhancement in
$Au+Au$ collisions. A measurement of $R_{AA}$ vs. system size may
be sensitive to the system size at which OZI becomes irrelevant to
$\phi$ production.

The $\phi$ $R_{CP}$ result is consistent with a partonic
recombination scenario \cite{das_hwa, fries, hwa2}.
In these models, the centrality dependence of the yield at
intermediate $p_t$ depends more strongly on the number of
constituent quarks than on the particle mass. Further higher
statistical data for the $\phi$ are needed to draw a conclusion.

In summary, STAR has measured $\phi$ meson production in
$\sqrt{s_{_{NN}}}=200$~GeV $Au+Au$ and NSD $p+p$ collisions at
RHIC. The $\phi/K^-$ yield ratios from $e^+e^-$, $p+p$ and $A+A$
collisions over a broad range of collision energy above the $\phi$
production threshold are remarkably close to each other. $\phi$
production, when scaled by the number of binary collisions, is
suppressed with respect to peripheral collisions in central
$Au+Au$ collisions.  The lack of a significant centrality
dependence of the $\phi/K^-$ ratio and the values of $\phi$
$\langle p_t \rangle$ effectively rule out kaon coalescence as a
dominant production channel for the $\phi$ at this energy.

We thank the RHIC Operations Group and RCF at BNL, and the NERSC
Center at LBNL for their support. This work was supported in part
by the HENP Divisions of the Office of Science of the U.S. DOE;
the U.S. NSF; the BMBF of Germany; IN2P3, RA, RPL, and EMN of
France; EPSRC of the United Kingdom; FAPESP of Brazil; the Russian
Ministry of Science and Technology; the Ministry of Education and
the NNSFC of China; SFOM of the Czech Republic, FOM and UU of the
Netherlands, DAE, DST, and CSIR of the Government of India; the
Swiss NSF.

%-------------------------------------------------------------------

\end{document}